\documentclass{sig-alternate}
\usepackage{mathptmx} 

\newcommand{\Name}{RecPipe}
\newcommand{\Infra}{RecPipe}
\newcommand{\Accel}{RPAccel}

\usepackage{enumitem}
\setlist[itemize]{noitemsep}
\usepackage{fancyhdr}
\usepackage[normalem]{ulem}
\usepackage[hyphens]{url}
\usepackage[sort,nocompress]{cite}
\usepackage[final]{microtype}
\usepackage[keeplastbox]{flushend}
\usepackage[bookmarks=true,breaklinks=true,colorlinks,linkcolor=black,citecolor=blue,urlcolor=black]{hyperref}

\fancypagestyle{firstpage}{
  \fancyhf{}
  
  \fancyfoot[C]{\thepage}
}

\title{Guidelines for Submission to MICRO 2021} 

\begin{document}
\title{ \Name: Co-designing Models and Hardware to Jointly Optimize Recommendation Quality and Performance}
\author{Udit Gupta$^{1,2}$, Samuel Hsia$^{1}$, Jeff (Jun) Zhang$^{1}$, Mark Wilkening$^{1}$, Javin Pombra$^{1}$, \\ Hsien-Hsin S. Lee$^{2}$, Gu-Yeon Wei$^{1}$, Carole-Jean Wu$^{2}$, David Brooks$^{1}$ \\ \\
$^{1}$Harvard University, $^{2}$Facebook AI Research \\ \\
ugupta@g.harvard.edu
}

\maketitle
\thispagestyle{firstpage}
\pagestyle{plain}


\begin{abstract}

Deep learning recommendation systems must provide high quality, personalized content under strict tail-latency targets and high system loads.
This paper presents RecPipe, a system to jointly optimize recommendation quality and inference performance.
Central to RecPipe is decomposing recommendation models into multi-stage pipelines to maintain quality while reducing compute complexity and exposing distinct parallelism opportunities.
RecPipe implements an inference scheduler to map multi-stage recommendation engines onto commodity, heterogeneous platforms (e.g., CPUs, GPUs).
While the hardware-aware scheduling improves ranking efficiency, the commodity platforms suffer from many limitations requiring specialized hardware.
Thus, we design RecPipeAccel (\Accel), a custom accelerator that jointly optimizes quality, tail-latency, and system throughput.
\Accel\ is designed specifically to exploit the distinct design space opened via RecPipe.
In particular, \Accel\ processes queries in sub-batches to pipeline recommendation stages, implements dual static and dynamic embedding caches, a set of top-$k$ filtering units, and a reconfigurable systolic array.
Compared to prior-art and at iso-quality, we demonstrate that \Accel\ improves latency and throughput by 3$\times$ and 6$\times$. 
\end{abstract}

\section{Introduction}

Deep neural network (DNN) based recommendation systems constitute an overwhelming fraction of AI cycles in production data centers (e.g., Facebook, Google, Alibaba)~\cite{hazelwood2018applied, gupta2020architectural, acun2020understanding, tpu, dinzhou2018deep, dienzhou2019deep, mtwnd, ncf, yiSysml18}. 
To improve content personalization in a wide range of services (e.g., search, e-commerce, movie and video-streaming, social media), the size of production recommendation models has grown by over 10$\times$ between 2017 and 2020~\cite{lui2020understanding,baidu_aibox,baidu_dist_inf}. 

\begin{figure}[t!]
  \centering
  \includegraphics[width=0.95\columnwidth]{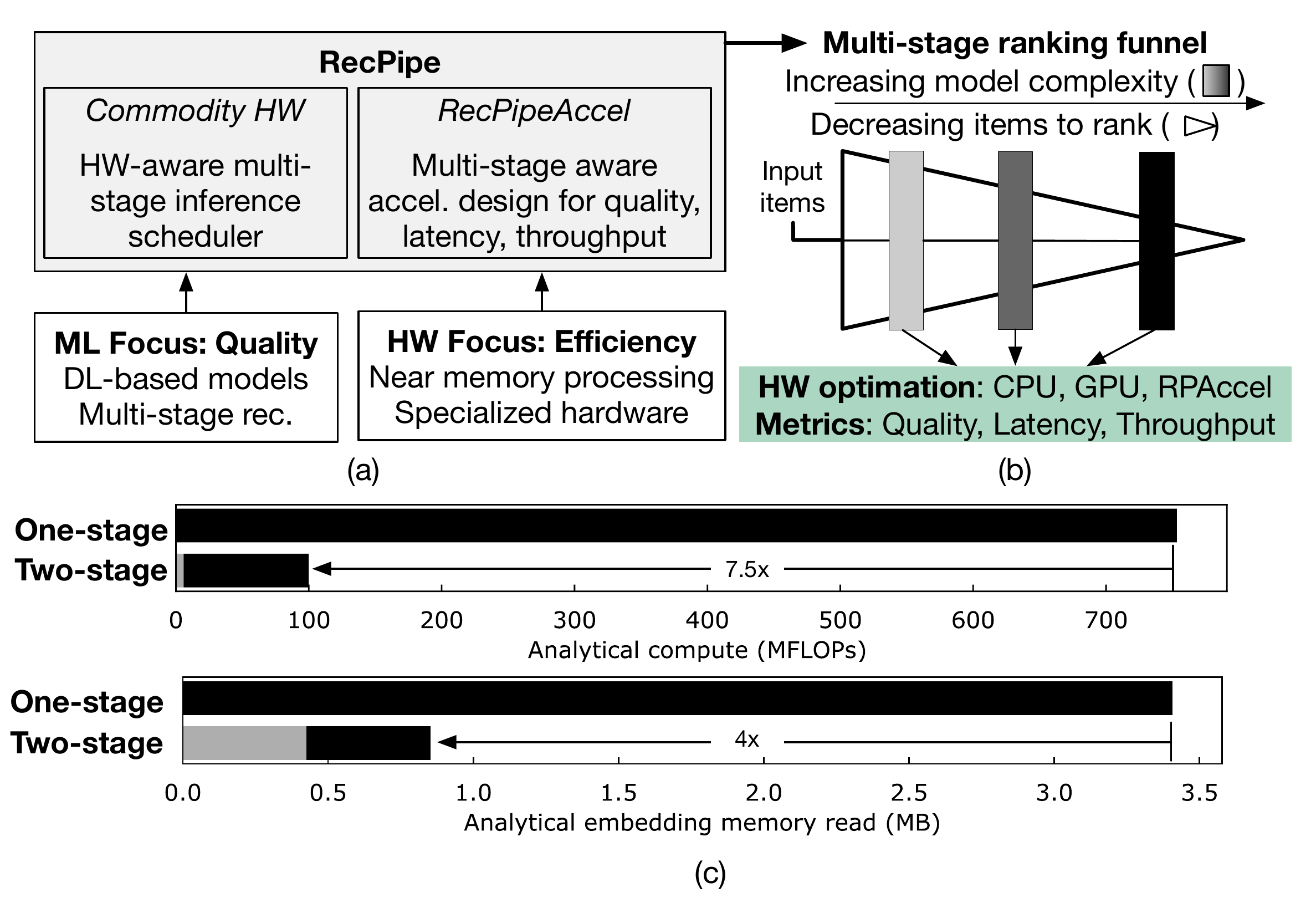}
  \vspace{-1.0em}
  \caption{ (a) Compared to prior work from machine learning and hardware researchers, this work jointly optimizes quality and performance. 
  (b) \Name\ co-designs models and hardware across multi-stage recommendation pipelines. 
  (c) Transforming monolithic models into multiple stages reduces overall compute demand and embedding memory accesses by 7.5$\times$ and 4.0$\times$, respectively.}
  \label{fig:overview}
  \vspace{-1em}     
\end{figure}


In response to the dramatic increase in infrastructure demands from the ever-increasing model complexity, system- and architecture-level solutions are customized for DNN-based recommendation, including inference schedulers~\cite{gupta2020deeprecsys}, near memory processing hardware~\cite{ke2020recnmp,kwon2019tensordimm,wilkening2021recssd}, and specialized accelerators~\cite{centaur, sambanova, jiang2021microrec}.
These prior solutions assume fixed models, leaving significant room for efficiency optimization. 
Evidenced by recent work optimizing DNNs for computer vision and natural language processing~\cite{eie, minerva, howard2017mobilenets, sandler2018mobilenetv2, epur, gupta2019masr, zhang2020model}, 
co-designing models with hardware is an effective approach.
However, the model accuracy requirement for recommendation tasks is stringent~\cite{dinzhou2018deep, dienzhou2019deep}, making the model-hardware co-design space challenging to navigate.

While accuracy represents a model's ability to predict whether users will like \textit{individual} items, production services are designed to serve users a personalized \textit{collection} of relevant items~\cite{jarvelin2002cumulated,chenRankingMetrics}. 
As such, while accuracy is intrinsic to models, \textit{quality} is optimized by improving model accuracy \textit{and} increasing the number of items ranked at the same time. 
The more holistic, application-level quality objective allows system architects to judiciously trade off accuracy for performance, opening new design spaces for system optimization.



Ranking all items with complex models is wasteful---only a small portion of items are relevant to individual users. Traditionally, recommendation engines achieve high quality by ranking a large number of input candidate items using complex DNNs.
The combination of large input working set sizes and complex models incurs high performance overheads.
Alternatively, one can decompose a monolithic ranking model into multiple stages to maintain overall quality at higher performance~\cite{instagramMsr, mtwnd, kang2019candidate}.
By splitting the monolithic model into two, a frontend model coarsely \textit{filters} relevant items while a more accurate backend model finely \textit{ranks} items to serve.
Further segmenting the pipeline into additional stages creates a ranking funnel (Figure~\ref{fig:overview} (b)) where complex models only rank items requiring accurate differentiation.
For the Criteo dataset and Facebook's Deep Learning Recommendation Model (DLRM)~\cite{criteo, naumov2019deep}, Figure~\ref{fig:overview}(c) shows that, at iso-quality,  compared to single-stage, multi-stage recommendation reduces memory and compute demands by 4.0$\times$ and 7.5$\times$, respectively.
This system-level view optimizing quality and efficiency motivates a new generation of hardware solutions for multi-stage recommendation.
Driven by this motivation, we propose {\it \Name}, a system to co-design recommendation models and hardware to improve both quality and performance (Figure~\ref{fig:overview}(a)).
Frontend stages pair light-weight models (e.g., low compute and memory demands) with large input sizes, exposing \textit{data-parallelism}.
Backend stages pair heavy-weight models (e.g., billions of FLOPs, many GBs of storage) with small input sizes, exposing \textit{model-parallelism} instead.
\Infra's system solutions exploit these distinct parallelism opportunities to jointly optimize quality, throughput, and tail-latency.

To understand the limits of commodity platforms, \Infra\ implements an inference scheduler that maps each recommendation stage across heterogeneous hardware (e.g., CPU, GPU) to maximize performance.
We find the optimal mapping depends on the application level targets and underlying hardware.
Despite the tight co-design between models and hardware, we find commodity CPU-GPU systems do not fully exploit the benefits of multi-stage recommendation as they suffer from low utilization and high PCIe communication overheads between stages.

To address these limitations, we design {\it RecPipeAccel} (\Accel), a specialized accelerator for multi-stage recommendation. 
Starting with a TPU-like, systolic array-based, recommendation accelerator~\cite{centaur}, \Accel's hardware optimizations improve efficiency at low area and power overheads. 
First, \Accel\ implements a reconfigurable systolic array that allows the hardware to concurrently process models across recommendation stages.
\Infra's inference scheduler provisions the fraction of systolic array resources to devote to frontend and backend stages based on application load, balancing latency and throughput.
Next, \Accel\ eliminates high PCIe communication overheads to the host processor by implementing multiple on-chip filtering units to identify the top-$k$ user-item interactions between stages.
Finally, to overlap frontend and backend query processing, \Accel\ breaks queries into sub-batches to pipeline stages and pre-fetch embedding vectors in separate caches.

The main contributions of this work include:
\begin{enumerate}[leftmargin=*]
\item We propose a new system, \Infra, that enables design space exploration and optimization for multi-stage recommendation inference. 
The framework integrates data sets (e.g., MovieLens~\cite{movielens1m}, Criteo~\cite{criteo}), models (e.g., neural matrix factorization~\cite{he2017neural}, DLRM~\cite{naumov2019deep}), and hardware (e.g., CPU, GPU, simulated accelerators) to study trade-offs among quality, tail-latency, and throughput.

\item We show designing and efficiently scheduling multi-stage pipelines for available commodity hardware platform reduces tail-latency by 4$\times$ and 3$\times$ on CPUs and heterogeneous CPU-GPU hardware, respectively.





\item We design \Accel, a novel accelerator that exploits the distinct properties of multi-stage recommendation to jointly optimize quality, latency, and throughput.
Compared to a state-of-the-art baseline accelerator~\cite{centaur}, \Accel's software and hardware optimizations reduce tail-latency by 3$\times$ and increases throughput by 6$\times$, at iso-quality as well as negligible area and power overheads.

\end{enumerate}

\begin{figure}[t!]
  \centering
  \includegraphics[width=\columnwidth]{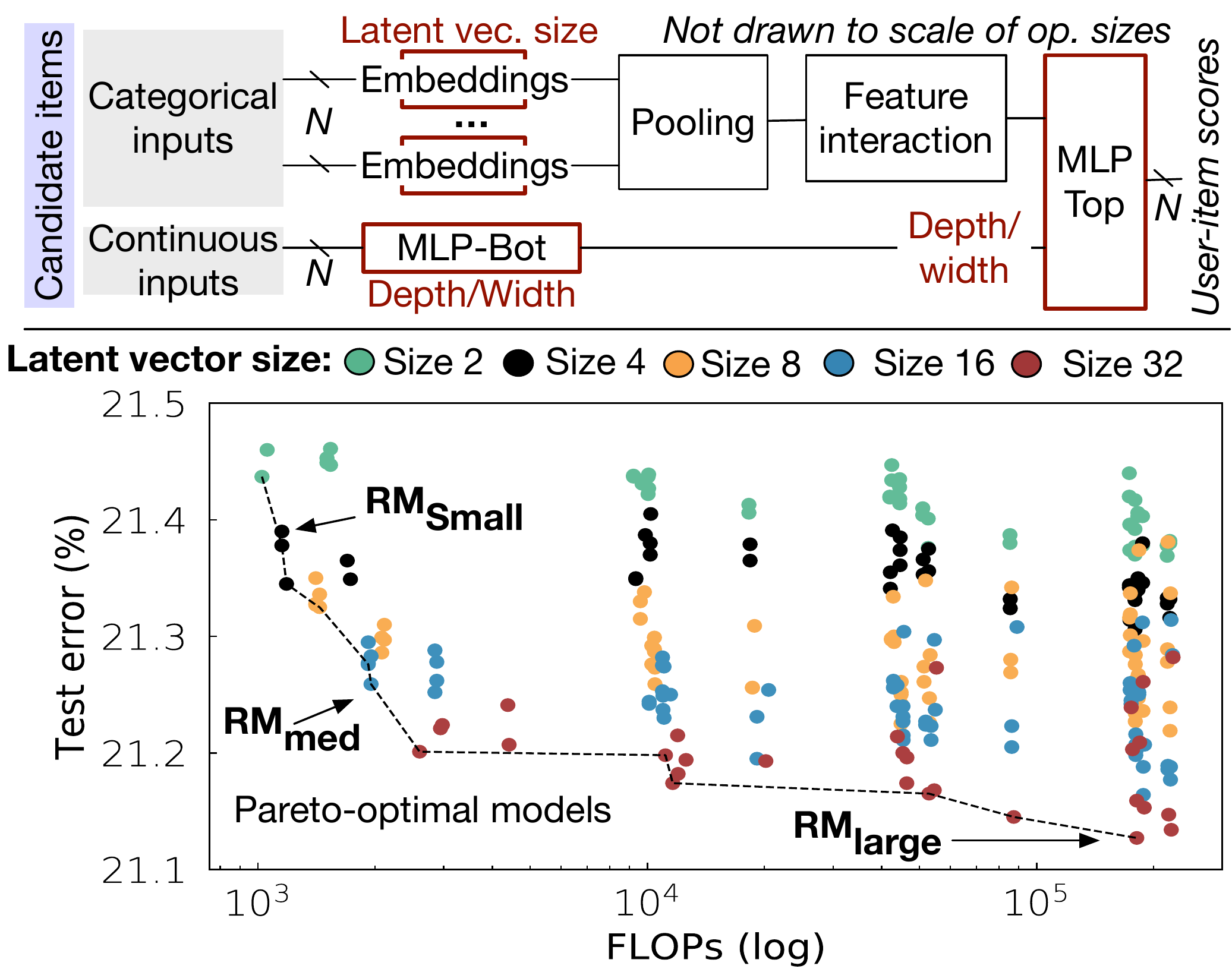}
  \vspace{-2em}
  \caption{(Top) General recommendation model architecture configured by embedding dimension and MLP size (outlined in red).  (Bottom) Hyperparameter sweep shows tradeoff between model complexity and error.}
  \label{fig:GeneralModel}
\end{figure}

\begin{table}[t!]
\begin{center}
\small
\begin{tabular}{|c||c|c|c|}
\hline
\textbf{Model} & \textbf{RM}$_{\textrm{small}}$ & \textbf{RM}$_{\textrm{med}}$ & \textbf{RM}$_{\textrm{large}}$ \\ \hline
Embedding Dim. & 4 & 16 & 32 \\ \hline
MLP-Bottom & 13-64-4 & 13-64-16 & 13-512-256-128-64-32 \\ \hline
MLP-Top & 64-1 & 64-1 & 96-1 \\ \hline
Model Size & 1GB & 4GB & 8GB \\ \hline
FLOPs & 1.1K & 2.0K & 180K \\ \hline
Model Error & 21.36\% & 21.26\% & 21.13\%\\ \hline
\end{tabular}
\end{center}
\vspace{-1.5em}
  \caption{ Pareto-optimal recommendation models. }
  \label{tab:models}
  \vspace{-1em}
\end{table}

\section{Motivation: Widening Design Space by Optimizing for Quality over Accuracy Alone}~\label{sec:quality}
Prior work on specialized systems for deep learning co-optimizes for model accuracy and run-time efficiency (performance, power, and energy)~\cite{eie, minerva, howard2017mobilenets, sandler2018mobilenetv2, epur, gupta2019masr}.
For neural recommendation however, hardware designers must go one step further, beyond accuracy, and optimize for quality. 
In this section, we first describe recommendation model architectures and conduct a model hyper-parameter sweep.
Then, we introduce the quality metric used in this work, showing the fundamental difference between accuracy and quality. 


\subsection{Training hyperparameter sweep}
Figure~\ref{fig:GeneralModel}(top) lays out the general architecture for DNN recommendation models~\cite{gupta2020architectural,naumov2019deep}. 
Continuous input features are processed with DNN layers, e.g. Multi Layer Perceptrons (MLP), while sparse input features are processed using embedding tables. 
Embedding tables are organized as a collection of embedding vectors with tens to hundreds of latent features.
Latent features map sparse inputs to low-dimensional, dense ones. 
By configuring the main network components (i.e., MLP depth/width, embedding latent vector dimension), highlighted in red, we realize models with varying storage capacity, compute demands, and accuracy. 

\begin{figure}[t!]
  \centering
  \includegraphics[width=\columnwidth]{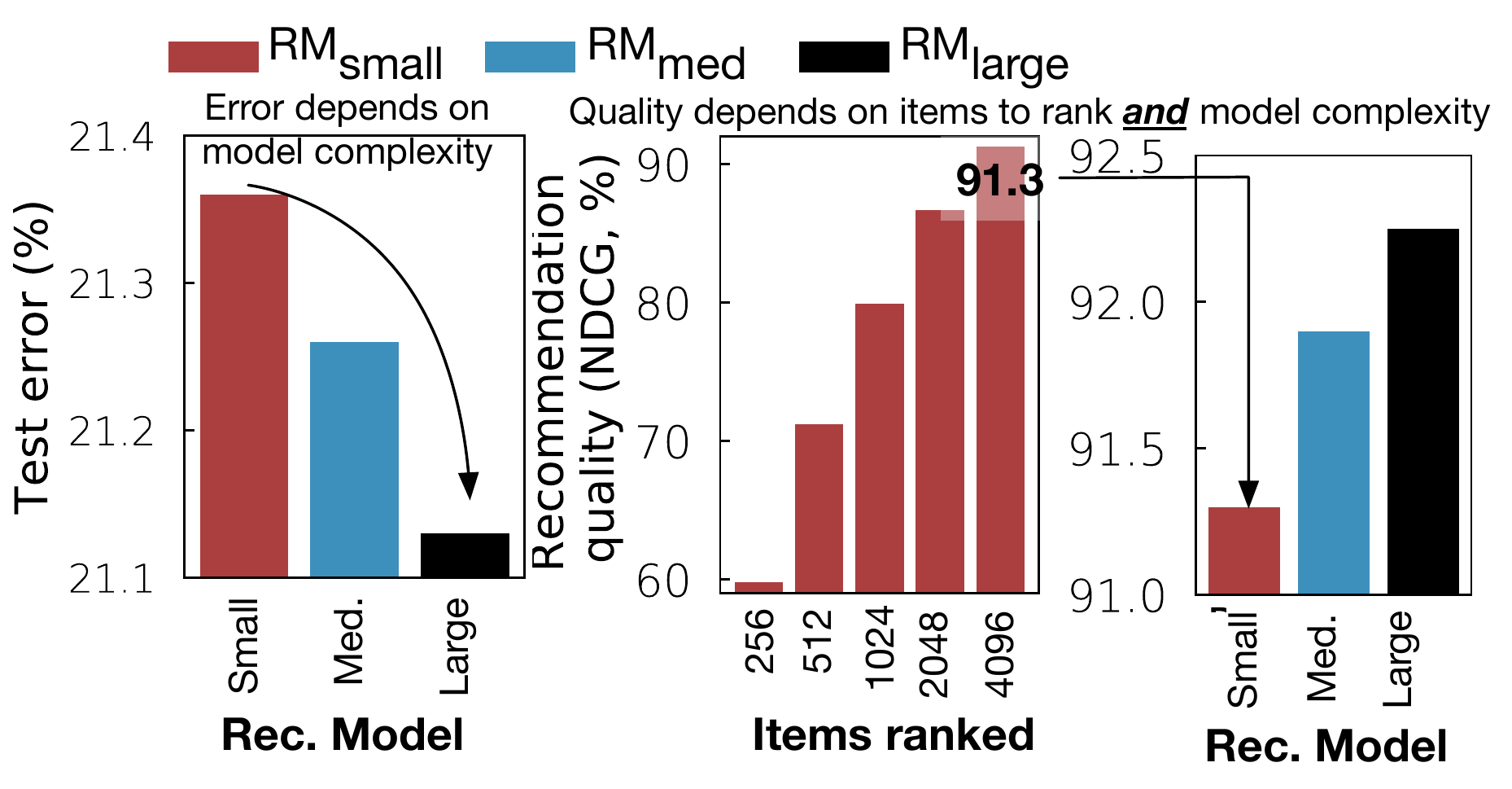}
  \vspace{-2.5em}
  \caption{ While accuracy depends only on model size (left), recommendation quality depends on number of items ranked (center) \textit{and} model size (right).}
  \label{fig:batching_motivation}
  \vspace{-1.5em}
\end{figure}

Figure~\ref{fig:GeneralModel}(bottom) shows a hyperparameter sweep by tuning the main network parameters for Facebook's DLRM trained on the Criteo dataset~\cite{naumov2019deep, criteo}.
Increasing the computational complexity of models reduces the test error. 
Models with 1.1K FLOPs and 180K FLOPs observe an error of 21.36\% and 21.13\%, respectively.
Note, a 0.23\% decrease in error is large given the high sparsity of user-item interactions in recommendation use cases~\cite{dinzhou2018deep, dienzhou2019deep}.
Recent industry publications show reductions of even 0.1\% error greatly improve user experience in real world applications~\cite{dinzhou2018deep, dienzhou2019deep}. 
Table~\ref{tab:models} shows the tradeoff between model error and complexity across three Pareto-optimal networks (i.e., RM$_{\textrm{small}}$, RM$_{\textrm{med}}$, RM$_{\textrm{large}}$).


\subsection{Quality versus accuracy} 
A model's accuracy represents its ability to correctly predict a user will positively interact with a \textit{single} item. 
However, in recommendation, models rank thousands of items opening the door for measuring overall quality. 
Quality measures the relevance of the \textit{entire collection} of items presented to users based on their personal preferences.
Following recent work from machine learning and recommendation systems researchers, we use normalized discounted cumulative gain (NDCG) to quantify the quality of the ordered list of output items.
%
NDCG~\cite{jarvelin2002cumulated,chenRankingMetrics} is the ratio between the measured and the ideal ordering, each of which is computed using discounted cumulative gain (DCG): for a list of $N$ items, $ DCG = \sum_i^N \frac{Rel_i}{log_2(i+1)} $.
$Rel_i$ represents item $i$'s score in the measured or ideal list and $log_2(i+1)$ discounts its relevance---dividing the score by the item's position in the list.


\textbf{Widening design space.} Compared to accuracy, optimizing for quality opens new system design opportunities.
For the Criteo dataset, Figure~\ref{fig:batching_motivation} illustrates the impact of varying the number of items ranked (x-axis) and model architecture (i.e., RM$_{\textrm{small}}$, RM$_{\textrm{med}}$, RM$_{\textrm{large}}$) on quality.
Empirically, the improvement in quality from increasing number of items ranked overshadows that from larger, more accurate models.
Note, the highest quality of 92.25 can be achieved by ranking all 4096 items with RM$_{\textrm{large}}$.
However, as shown in Figure~\ref{fig:overview}, decomposing monolithic models into multiple stages, where small models filter relevant items and large models perform fine-grained ranking, improves computational efficiency at iso-quality.
At the frontend, candidate items are coarsely ranked with models that incur memory and compute demands.
This reduces the list of candidate items (i.e., working set size) incrementally over the stages. 
Subsequent stages use larger models for finer-grained ranking.
Going beyond accuracy, quality depends on the number of stages, network architectures, and the number of items ranked at run-time: widening the design space to co-optimize performance and quality.

\begin{figure*}[t!]
  \centering
  \includegraphics[width=\textwidth]{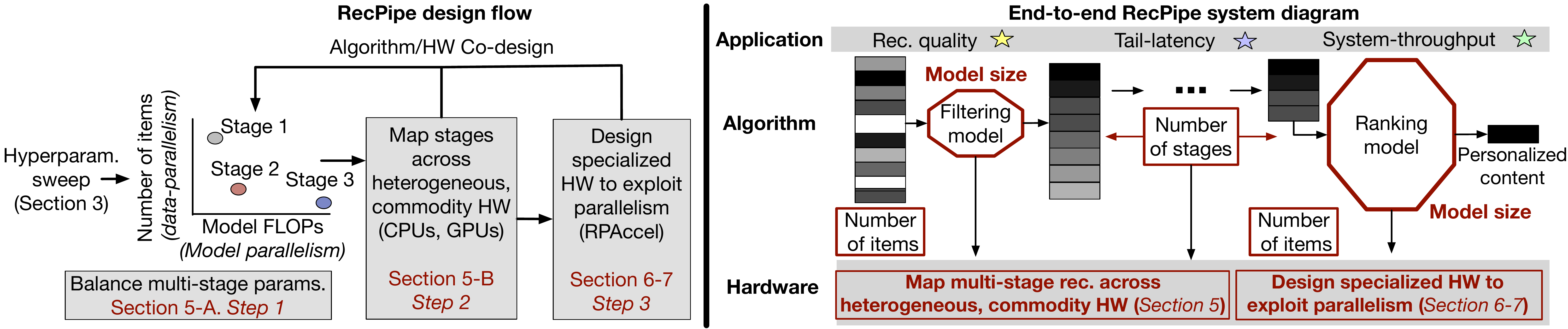}
  \vspace{-2em}
\caption{The structure of a multi-stage recommendation pipeline. Highlighted in red, \Infra~explores a variety of recommendation model and hardware infrastructure parameters to balance quality, latency, and throughput.
}
  \label{fig:MSRSystem}
  \vspace{-1em}
\end{figure*}

\begin{figure*}[t!]
  \centering
  \includegraphics[width=\textwidth]{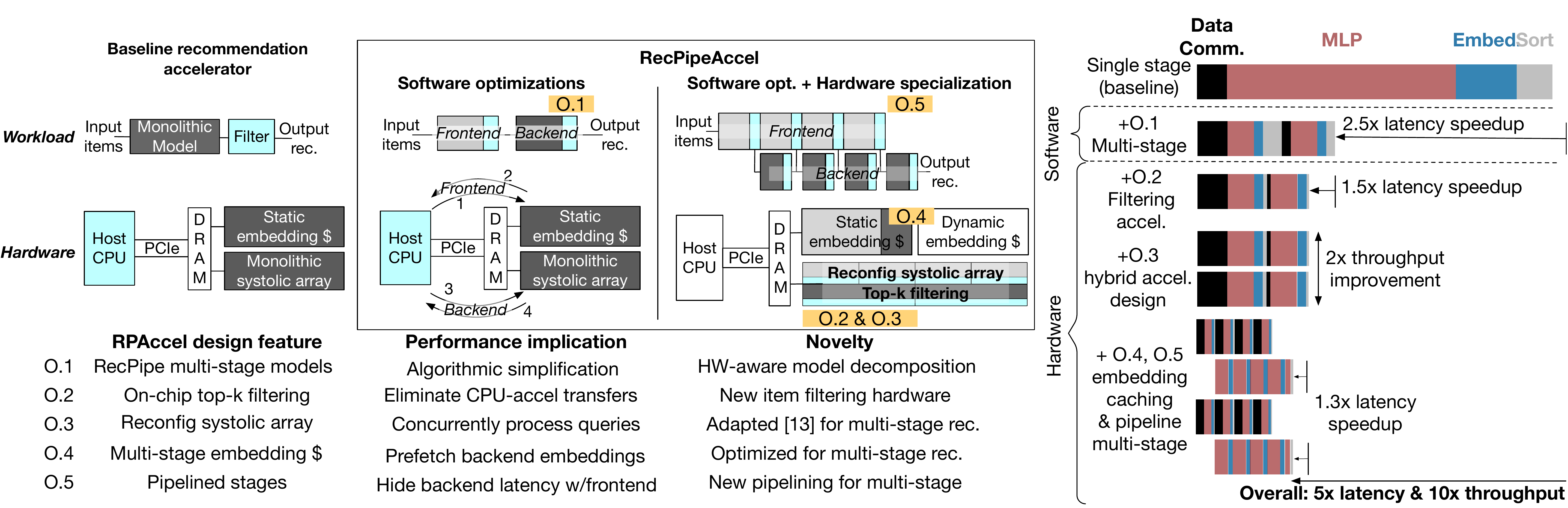}
  \vspace{-2em}
\caption{Comparison of baseline recommendation hardware accelerator and \Accel.
We describe \Accel's five main innovations (i.e., O.1 to O.5) on the left and their performance benefits in the ablation study on the right.}
  \label{fig:RPAccelSummary}
  \vspace{-1em}
\end{figure*}

\section{\Infra\ Design: A System to Optimize Multi-Stage Recommendation Inference}~\label{sec:flow}
We propose \Infra, a novel system to explore the model- and hardware-level design space to collectively optimize recommendation quality, tail-latency, and system throughput.
Figure~\ref{fig:MSRSystem}(left) shows \Infra's multi-step design process.
The input to \Infra~is a Pareto-frontier of recommendation models balancing model accuracy and complexity.
To co-optimize quality and hardware efficiency on commodity platforms, \Infra\ balances multi-stage parameters and statically schedules each stage across available hardware resources (i.e., CPUs and GPUs).
Going further, \Infra~exposes distinct parallelism opportunities that are exploited by designing specialized hardware.  
Figure~\ref{fig:MSRSystem}(right) illustrates the multi-stage recommendation pipeline and the design space optimized by \Infra.
The model-level and hardware-level design parameters are highlighted in red. 
We detail how \Infra~co-designs these parameters to maximize quality and performance below.

\subsection{Hardware-aware multi-stage scheduling} 
\Infra\ implements a post-training, inference scheduler customized for multi-stage recommendation.
In step 1, \Infra\ balances multi-stage modeling parameters.
In step 2, \Infra\ exploits the parallelism opportunities exposed from step 1, and maps stages across heterogeneous hardware.

\textbf{Algorithmic scaling (Step 1).} \Infra~exhaustively explores the design space of pairing Pareto-optimal recommendation models and number of items to rank at each stage in the multi-stage pipeline.
In the frontend, lightweight models are paired with large working set sizes exhibiting high data-level parallelism; in the backend heavyweight models are paired with smaller working set sizes exhibiting high model-level parallelism.
By collectively balancing model complexity and input working set size, \Infra~maximizes overall quality under strict latency targets and system loads. 


\textbf{Heterogeneous hardware mapping (Step 2).} 
Given the distinct parallelism opportunities from the aforementioned algorithmic scaling step, \Infra~exhaustively explores the mapping of multi-stage models on available hardware at the stage granularity.
We begin by considering commodity hardware platforms i.e., CPUs and GPUs.
GPUs implement a highly data-parallel architecture that parallelize individual queries, especially in the frontend with large working set sizes.
On the other hand,  many-core CPUs can simultaneously process multiple queries providing high-throughput.
\Infra~exploits these architectural differences to schedule each recommendation stage onto the underlying hardware.
In fact, we find the optimal mapping of multi-stage recommendation varies across application-level targets (e.g., tail-latency, system load).
Thus, \Infra~schedules multi-stage pipelines onto available hardware, based on algorithmic model parameters, architectural characteristics, and application-level requirements, to maximize quality and performance.


While achieving the maximal quality target and at iso-throughput, the scheduling optimizations reduce tail-latency by 4$\times$ on CPUs and 3$\times$ on heterogeneous hardware i.e., CPUs and GPUs (see Section~\ref{sec:cpu} for details).
However, despite these performance improvements there remains significant room for further optimization.
In particular, the commodity CPU-GPU platforms suffer from two main drawbacks.
First, GPUs exhibit low utilization when exploiting data-level parallelism in the frontend and model-level parallelism in the backend, primarily due to the high overhead of embedding lookups and memory transformation operations on GPUs~\cite{gupta2020deeprecsys}.
Second, between stages, high PCIe communication overheads across the CPU and GPU limit achievable throughput.
To address these limitations, and given the importance of data center-scale recommendation, \Infra~enables designing specialized hardware for multi-stage recommendation.

\subsection{Custom hardware to accelerate multi-stage recommendation}~\label{sec:accel_summary}
Figure~\ref{fig:RPAccelSummary} illustrates the high-level architecture of the proposed recommendation accelerator, \Accel.
On the left, we start with a state-of-the-art accelerator baseline that minimizes inference latency for a single-stage recommendation model using a TPU-like monolithic systolic array and static cache for \textit{hot-embeddings}~\cite{centaur}.
The aforementioned software optimizations reduce workload complexity by decomposing the single-stage model into a multi-stage pipeline.
Given the simplified workload, \Accel\ is designed to concurrently process multiple models and queries, end-to-end.
Figure~\ref{fig:RPAccelSummary}(right) provides an ablation study for the proposed software and hardware optimizations, demonstrating significant latency and throughput improvement potential (i.e., O.1 to O.5).



By exploiting unique properties of multi-stage recommendation, \Accel~is designed to balance both inference latency and throughput based on application-level requirements.
\begin{itemize}[leftmargin=*]
\item (O.1) \Infra~decomposes a single-stage model into multiple stages (2.5$\times$ latency reduction).
\item (O.2) \Accel~comprises a top-$k$ filtering unit to identify the $k$ highest quality items based on predicted click-through-rate (CTR) to be ranked by subsequent stages; this eliminates host-accelerator communication between recommendation stages (1.5$\times$ latency reduction).
\item (O.3) \Accel~implements a reconfigurable systolic array to concurrently process multiple stages and queries (2$\times$ hardware utilization and throughput). \Infra's software scheduler (see above) splits the monolithic systolic array into multiple sub-arrays based on application-level targets (quality, latency, throughput) and multi-stage models.
\item (O.4) \Accel~balances on-chip memory resources to statically cache \textit{hot-embeddings} and dynamically prefetch embeddings for backend models (40\% reduction in average memory access time). The static cache is provisioned for both frontend and backend stages; the dynamic cache prefetches embeddings for the backend as the frontend finishes sub-batches of the input query.
\item (O.5) \Accel~breaks queries into sub-batches to pipeline -- and thus -- overlap computation from frontend and backend stages (1.3$\times$ latency reduction).
\end{itemize}
While achieving the highest quality target, compared to the baseline recommendation accelerator, \Accel's optimizations collectively decrease tail-latency by up to 5$\times$ and increase throughput by up to 10$\times$ (see Section~\ref{sec:accel}-\ref{sec:accel_eval} for details).

\begin{figure}[t!]
  \centering
  \includegraphics[width=\columnwidth]{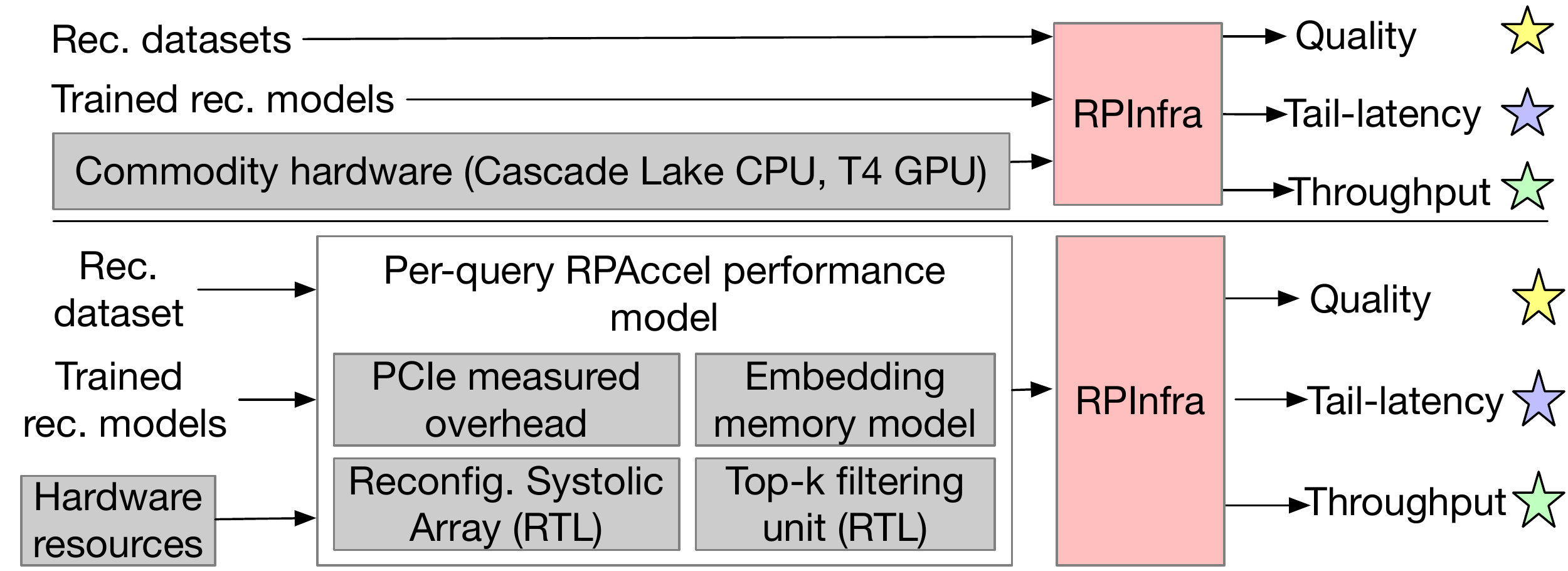}
  \vspace{-1.5em}
\caption{Evaluation methodology of \Infra~on commodity (top) and specialized (bottom) hardware.}
  \label{fig:method}
\end{figure}

\begin{table}[t!]
\begin{center}
\small
\begin{tabular}{|c||c|c|c|}
\hline
\textbf{Machines} & \textbf{Cascade Lake CPU} & \textbf{NVIDIA T4 GPU} \\ \hline
Frequency & 2.8 GHz &  585 MHz \\ \hline
Cores & 64 & 2560 \\ \hline
SIMD & AVX-512 & FP32x64\\ \hline
Cache Sizes & 1-16-22 MB & 96-512 KB\\ \hline
DRAM Capacity & 384 GB & 15 GB\\ \hline
DDR Bandwidth & 75 GB/s & 300 GB/s \\ \hline
TDP & 300 Watt & 70 Watt \\ \hline
\end{tabular}
\end{center}
\vspace{-1em}
  \caption{ Commodity hardware in experimental setup. }
  \label{tab:machines}
  \vspace{-1em}
\end{table}

\section{Experimental Methodology}~\label{sec:method}
Figure~\ref{fig:method} illustrates the evaluation methodology we use to study the system design implications of multi-stage recommendation.
\Infra~encompasses a vast design space across multi-stage modeling parameters, hardware solutions, and application-level targets.
To foster deeper understanding, we analyze cross-sections of the design space based on the application-level targets: iso-quality, iso-throughput, and iso-latency. 
This section details the methodology on both real, commodity hardware and simulated, specialized hardware.

\textbf{Datasets and models.} 
We evaluate \Infra\ with three open-source datasets: Criteo Kaggle~\cite{criteo}, MovieLens 1M~\cite{movielens1m}, and MovieLens 20M~\cite{movielens1m}.
We train neural matrix factorization models for both MovieLens datasets~\cite{ncf}.
To provide intuition across the large design space studied in this work, we conduct a deep dive using Criteo and Facebook's DLRM~\cite{naumov2019deep}.
On top of this deep dive, Section~\ref{sec:summary} summarizes results across all datasets.
All models are implemented in PyTorch.

\textbf{Application-level targets.}
This work optimizes recommendation based on three application-level targets:

\begin{itemize}[leftmargin=*]
\item Quality: We use NDCG~\cite{jarvelin2002cumulated,chenRankingMetrics} to quantify recommendation quality of the top sixty-four items served.
For commensurate analysis, final results are presented based on the highest quality achieved for each model and dataset: NDCG of 92.25 for Criteo (see Section~\ref{sec:quality}).

\item Tail-latency: To maintain user-experience, recommendations must meet SLAs and be served under strict tail-latency targets~\cite{gupta2020deeprecsys}, measured as \textit{99}$^{th}$ percentile (\textit{p99}).

\item Throughput: Data-center recommendation systems must maximize throughput, measured as the queries processed per second (QPS). Queries  follow a Poisson arrival rate.
\end{itemize}


\textbf{Commodity hardware systems.} 
To study the proposed designs in the context of data center scale recommendation, \Infra~runs datasets and models directly on real CPUs (server class Intel Cascade Lake) and GPUs (NVIDIA T4).
Refer to Table~\ref{tab:machines} for detailed hardware specifications.
Experiments on CPUs use multiple processes to exploit parallelism across cores---each core has a single PyTorch/MKL thread.
GPUs use CUDA/cuDNN 10.1.

\begin{table}[t!]
\begin{center}
\small
\begin{tabular}{|c||c|}
\hline
\textbf{Parameter} & \textbf{\Accel~configuration} \\ \hline
Frequency & 250 MHz \\ \hline
Systolic Array SRAM capacity & 8MB \\ \hline
Systolic Array MAC units & 128$\times$128 MACs \\ \hline
Embedding cache capacity & 16MB \\ \hline
DRAM capacity & 16 GB \\ \hline
DRAM bandwidth & 64 GB/s \\ \hline
DRAM latency & 100 cycles \\ \hline
\end{tabular}
\end{center}
\vspace{-1.5em}
  \caption{Fixed resources in \Accel. }
  \label{tab:accel}
  \vspace{-1em}
\end{table}


\begin{figure*}[t!]
  \centering
  \includegraphics[width=\textwidth]{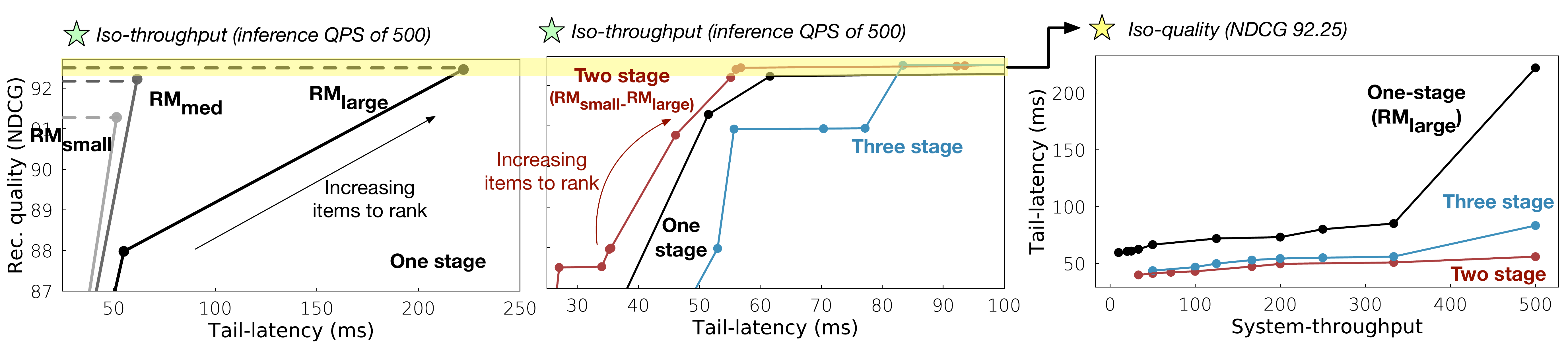}
  \vspace{-2em}
  \caption{ (Left) In single-stage recommendation, larger models achieve the higher quality at the expense of tail-latency.
  (Middle) Tuning multi-stage parameters improves quality under strict performance constraints.
  (Right) While achieving the highest-quality target, decomposing single-stage recommendation to multiple stages reduces tail-latency.}
  \label{fig:msr_cpu}
  \vspace{-1em}
\end{figure*}

\textbf{Accelerator modeling.}
\Infra~uses a two-step evaluation methodology to simulate specialized hardware.

First, we evaluate the latency of each query across each stage of the multi-stage pipeline.
The latency per stage is computed as cumulative time from data transfers over PCIe, embedding lookups, MLP operations, and the top-$k$ filtering units. 
Host-to-accelerator PCIe overheads are based on real measurements from the CPU-GPU system (see Table~\ref{tab:machines}).
For embedding lookups, we compute hit rates based on the cache locality of open-source datasets.
Given the cache hit rates, we compute the memory latency of embedding operations using simple latency and bandwidth models for SRAM and DRAM.
For MLP layers, We design and implement the systolic array and the top-$k$ filtering unit in RTL to gather cycle-accurate performance measurements, including overheads from loading weights and activations from DRAM.
Combining latency for all stages forms per-query performance model.

Second, the per-query latencies are fed into \Infra~which simulates the at-scale performance characteristics of \Accel, measuring tail-latency, system-throughput, and quality, of processing tens of thousands of queries.

For area and power evaluations, we separately synthesize the reconfigurable systolic array, top-$k$ filtering unit, and memories in a 12$nm$ FinFET technology. 
As shown in Table~\ref{tab:accel}, \Accel\ implements comparable compute and memory resources to a data center TPU accelerator (40 Watt TDP)~\cite{tpu}. 



\section{Evaluation of RecPipe Inference Scheduler on Commodity Hardware}~\label{sec:cpu}
In this section we use \Infra~to efficiently schedule multi-stage recommendation onto heterogeneous hardware available in data centers. 
First, \Infra\ balances the multi-stage modeling parameters---number of stages, models per stage, items to rank per stage---to co-optimize tail-latency, throughput, and quality.
Next, \Infra\ co-designs the multi-stage parameters for heterogeneous systems comprising CPUs and GPUs. 
We show the optimal configuration of multi-stage parameters depends on the underlying hardware.
Furthermore, we show that while GPUs enable higher throughput and quality at low-latency targets, CPU-only execution achieves higher throughput under more relaxed latency targets.

\subsection{Mapping multi-stage pipelines to CPUs}~\label{sec:cpueval}
Figure~\ref{fig:msr_cpu}(left) illustrates the tradeoff between tail-latency and quality for single-stage recommendation on CPUs.
Following intuition, larger more complex models (e.g., RM$_{\textrm{large}}$) achieve higher quality at the expense of higher tail-latency.

\textbf{Takeaway 1:} \textit{Carefully balancing multi-stage parameters unlocks higher recommendation quality and throughput at strict tail-latency targets.}

At a fixed system load (i.e., QPS of 500), Figure~\ref{fig:msr_cpu}(center) shows tradeoff between tail-latency and quality for one-, two-, and three-stage designs.
Exhaustively sweeping all possible combinations of models per stage and number of items to rank per stage, we show the Pareto-frontier results. 


Compared to single-stage designs, Figure~\ref{fig:msr_cpu}(center) shows multi-stage designs achieve higher quality under strict performance constraints.
The single-stage design ranks all 4096 items with RM$_{\textrm{large}}$.
The optimal two-stage design first processes 4096 items with RM$_{\textrm{small}}$ followed by the top 256 items with RM$_{\textrm{large}}$, reducing tail-latency by 4$\times$ given the lower compute and memory demands.

The importance of optimizing for quality, not accuracy, can be seen by diving deeper into the two-stage design.
To achieve high quality, the backend implements the most accurate network (i.e., RM$_{\textrm{large}}$); the frontend implements either RM$_{\textrm{med}}$ or RM$_{\textrm{small}}$.
While RM$_{\textrm{med}}$ has higher accuracy, the benefits are overshadowed by the additional compute and memory requirements (see Table~\ref{tab:models}).
In fact, with RM$_{\textrm{large}}$ in the backend, both frontend options achieve the same quality (NDCG 92.25).
But, the combination of RM$_{\textrm{med}}$-RM$_{\textrm{large}}$ has a 1.6$\times$ longer tail-latency compared to RM$_{\textrm{small}}$-RM$_{\textrm{large}}$.
Designers must jointly optimize for quality and performance.

In addition to quality, balancing multi-stage parameters improves throughput at strict tail-latency targets.
Figure~\ref{fig:msr_cpu}(right) shows the tradeoff between tail-latency and throughput, at the highest quality target (NDCG of 92.25).
Compared with the one-stage system, the two-stage pipeline reduces tail-latency by 4.4$\times$ (QPS of 500).
However, decomposing the pipeline into three stages decreases performance given additional queuing delays between stages, which overshadow the 30\% reduction in compute  between two- and three-stage designs.
Note, the tradeoffs will vary across datasets---varying model complexities and items to rank per stage will impact the optimal configuration (see Section~\ref{sec:summary} for examples). 
\subsection{Mapping multi-stage pipelines to heterogeneous systems}~\label{sec:gpu}
\begin{figure}[t!]
  \centering
  \includegraphics[width=\columnwidth]{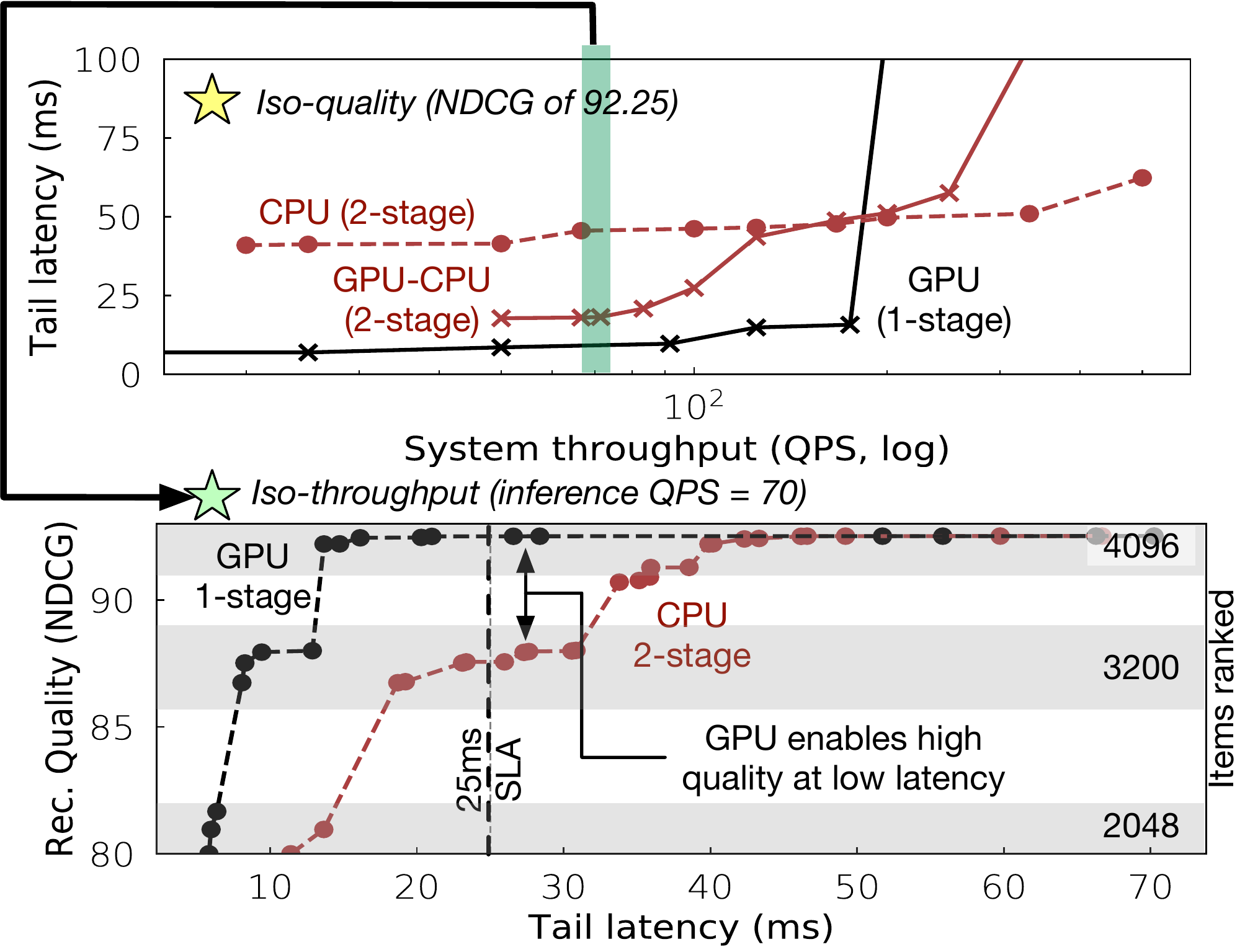}
  \vspace{-2.0em}
  \caption{ (Top) At iso-quality, mapping frontend (i.e., data-parallel) stages to GPUs reduces tail-latency by up to 3$\times$; CPU-only execution achieves higher system throughput.
  (Bottom) At a lower system throughput (i.e., QPS of 70), the lower latency on GPUs can be traded off for higher quality compared to CPU-based execution.}
  \label{fig:gpu_cpu_iso_acc}
  \vspace{-1em}
\end{figure}

Figure~\ref{fig:gpu_cpu_iso_acc}(top) illustrates the tradeoff between throughput and tail-latency while achieving the high quality target (NDCG of 92.25). 
Using \Infra, we exhaustively evaluate all mappings between multi-stage recommendation and heterogeneous hardware and show the best configurations: one-stage GPU-only, two-stage GPU-CPU, and the two-stage CPU-only configurations in Figure~\ref{fig:gpu_cpu_iso_acc}(top).
For the two-stage GPU-CPU design, \Infra~maps either the frontend or the backend to the GPU, running the other on the CPU.
In particular, we show results for frontend running on the GPU and backend on the CPU as our empirical evaluations show it provides higher performance. 
We also evaluate mapping two stages to the GPU with multi-tenant execution. 
Our evaluations show this configuration is unable to extract the fine-grain parallelism from  multi-stage's data dependency, incur longer latency than the one-stage GPU-only configuration.

\textbf{Takeaway 2:} \textit{Given architectural differences, the optimal multi-stage parameters vary on CPUs versus GPUs.}

Recall from our previous analysis, for CPU-only execution the two-stage design achieves the highest performance;  on the heterogeneous system, however, the single-stage GPU-only configuration (solid black) achieves higher performance than multi-stage using both CPU and GPU (solid red). 
The reason is twofold.
First, we observe comparable latency for RM$_{\textrm{small}}$ versus RM$_{\textrm{large}}$ on the GPU, overshadowing the benefits of decomposing models into finer-grained pipelines.
Second, the multi-stage GPU-CPU design requires transferring more intermediate results across PCIe, incurring heavy queuing delays and limiting system performance.

Nonetheless, the multi-stage GPU-CPU design plays an important role. 
Recent work shows production-scale recommendation model sizes are growing rapidly---by an order of magnitude in just three years~\cite{lui2020understanding}.
For production-scale models that are larger than the DRAM capacity available on GPUs (e.g., $\sim$ 15GB on NVIDIA T4), designers will need to decompose models into multiple stages.
Here, frontend stages run on the GPU in order to circumvent storage capacity limits and exploit data-parallelism with the larger input working set size; the backend models run on the CPU.
Figure~\ref{fig:gpu_cpu_iso_acc}(top) shows that this multi-stage GPU-CPU design achieves up to 3$\times$ lower latency than the multi-stage CPU-only configuration.


\begin{figure*}[t!]
  \centering
  \includegraphics[width=\textwidth]{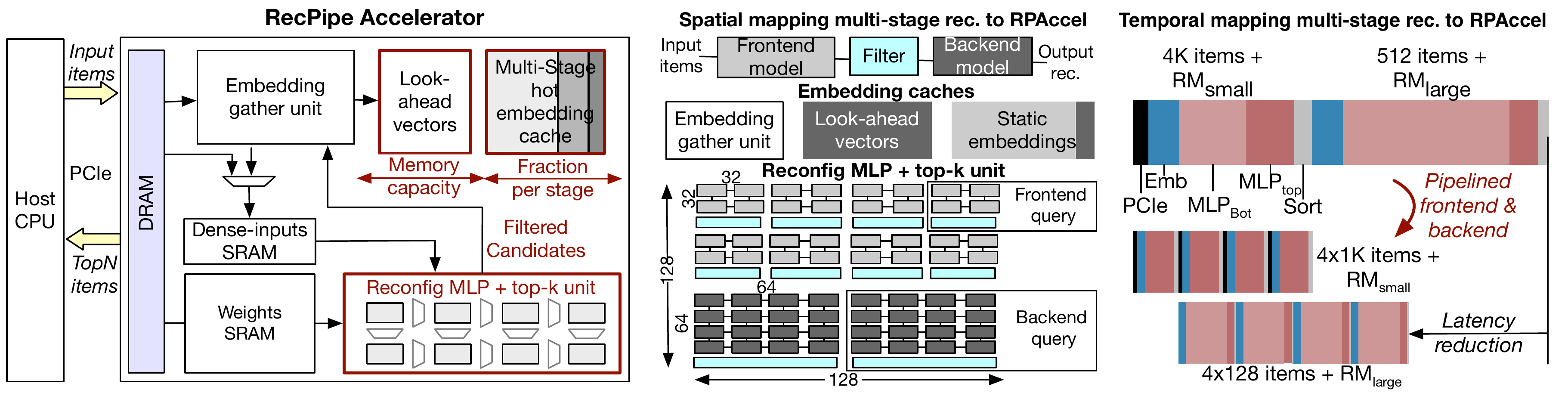}
  \vspace{-2.5em}
  \caption{(Left) Overall design of the RecPipe accelerator (\Accel) comprising an embedding gather unit with two on-chip caches for static and dynamic vectors, and a reconfigurable MLP and top-k filtering unit.
  (Middle) Static mapping of multi-stage recommendation onto \Accel. Frontend and backend share both memory and compute resources.
  (Right) Temporal mapping of multi-stage recommendation onto \Accel~with pipelined frontend and backend models.
    }
  \label{fig:MSRAccel}
  \vspace{-0.5em}
\end{figure*}

\begin{figure*}[h]
  \centering
  \includegraphics[width=\textwidth]{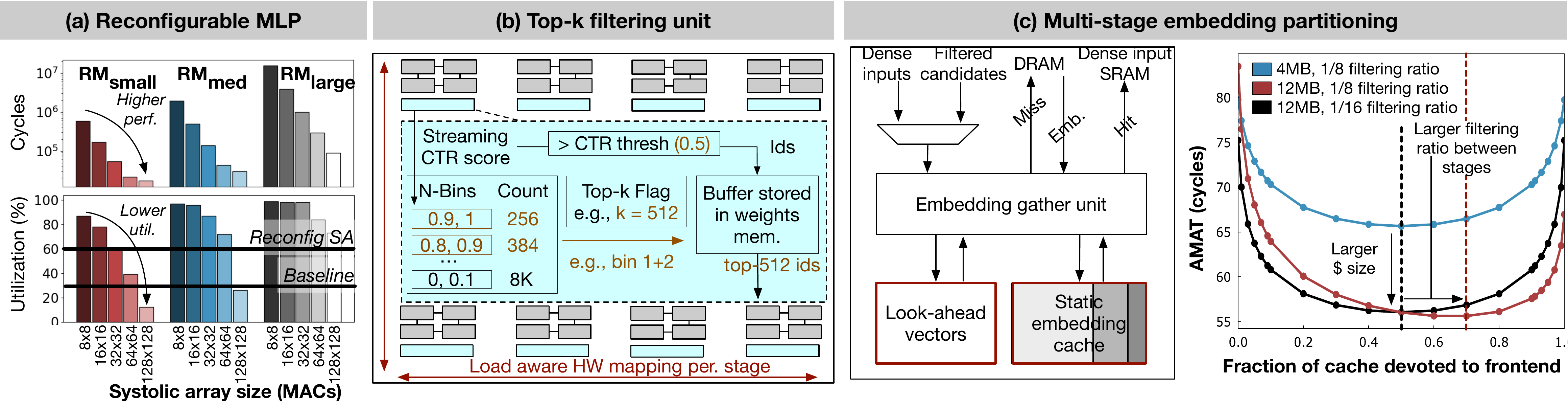}
  \vspace{-2.0em}
  \caption{ Design space exploration of \Accel. (a) Larger systolic arrays suffer from low utilization on smaller models, motivating provisioning resources into sub-arrays for concurrent query processing. Compared to a monolithic array with 30\% utilization, the reconfigurable array has a 60\% utilization.
  (b) Top-$k$ filtering unit designed to minimize area and power while eliminating host-accelerator PCIe communication overheads.
  (c) On-chip embedding cache resources must be asymmetrically provisioned across frontend and backend to minimize average memory access time. }
  \label{fig:rpaccel_dse}
  \vspace{-1em}
\end{figure*}

\textbf{Takeaway 3:} \textit{By maximizing throughput at low latency, GPUs unlock higher recommendation quality.}

Despite the GPUs achieving 3$\times$ lower latency than the CPU-only designs (see Figure~\ref{fig:gpu_cpu_iso_acc}(top)), the GPUs remain underutilized with an occupancy of 25\%, and memory and power utilization of 10\% and 45\%, respectively.
Improving utilization requires higher batching.
Unfortunately, as we increase batching and system throughput (x-axis), the GPU-enabled designs suffer from a sudden degradation in tail-latency due to high queuing delays; in comparison, the CPUs sustain higher throughput by concurrently processing queries across cores (e.g., task-parallelism).

While the latency reduction from GPU's does not translate to higher throughput, it can enable higher quality.
Figure~\ref{fig:gpu_cpu_iso_acc}(bottom) illustrates the tradeoff between tail-latency and quality for CPU- and GPU-based recommendation at iso-throughput.
Following our previous results, we show the optimal configurations: single-stage GPU-only and two-stage CPU-only designs.
Given the fixed models, \Infra~tradeoffs off latency for quality by increasing the number of items ranked per query.
At a strict SLA target of 25$ms$, the CPU achieves an NDCG of 87, while the GPU achieves an NDCG of 92.25. 
The increase in quality is a direct result of GPU's data-parallel architecture allowing it to rank 4096 items compared to the CPU ranking only 3200 items at the 25 $ms$ SLA.
Thus, AI accelerators for recommendation must be evaluated not only for performance benefits but also on quality achieved under strict performance and resource constraints.


\textbf{Limitations of commodity hardware.}
Based on the performance analysis above, we identify multiple limitations of commodity platforms running multi-stage recommendation. 
In particular, GPUs do not directly benefit from decomposing models into multiple stages. 
This is due to the limits of multi-tenant execution, under utilized hardware when separately exploiting data- and model- level parallelism across stages, and high PCIe data communication between stages. 
Given these limitations and the growing scale of personalized recommendation across Internet services~\cite{lui2020understanding, dienzhou2019deep, dinzhou2018deep}, we use \Infra\ to unlock the opportunities from multi-stage ranking by designing specialized hardware to provide high quality and infrastructure efficiency, in the following section.

\section{Analysis of RecPipeAccel's Design Space}~\label{sec:accel}
This section proposes \Accel,~a specialized accelerator tailored to multi-stage recommendation models.
We start with a baseline TPU-like recommendation accelerator~\cite{centaur}.
The baseline optimizes for low-latency single-stage inference, but suffers from low utilization and system throughput on multi-stage pipelines.
To accelerate multi-stage recommendation, as summarized in Section~\ref{sec:accel_summary}, \Accel~comprises four main features that exploit distinct opportunities enabled by \Name: the pipeline execution, a reconfigurable MLP unit, a top-$k$ filtering unit, and the partitioned embedding cache for hot-vectors across models and prefetched backend vectors.

\subsection{Mapping multi-stage pipelines to \Accel}
Figure~\ref{fig:MSRAccel}(left) illustrates the high-level architecture of \Accel.
Unlike prior art which accelerates \textit{single-stage} model inferences alone, \Accel\ is designed to process queries end-to-end: model inferences for  \textit{multiple stages} and \textit{filtering} top-$k$ user-item interactions between stages. 
Figure~\ref{fig:MSRAccel}(center) shows how multi-stage recommendation is mapped onto \Accel.
Networks across the stages share accelerator memory and compute resources.
For each stage, to produce predicted CTR scores for each user-item pair, \Accel~comprises an MLP and embedding gather unit.
\Accel\ implements a set of top-$k$ filtering units to identify high-quality user-item pairs.




\textbf{Takeaway 4:} \textit{Breaking queries into multiple sub-batches enables pipelined execution of frontend and backend stages.}

Figure~\ref{fig:MSRAccel}(right) shows the temporal mapping of multi-stage recommendation onto \Accel.
To reduce latency, \Accel~pipelines frontend and backend stages by breaking queries into smaller sub-batches.
As an example, Figure~\ref{fig:MSRAccel}(right) shows \Accel~splitting a single query of 4K items into four smaller batches of 1K each, overlapping frontend and backend stages.
The degree of sub-batching must be carefully balanced in order to maintain high utilization and quality.
Smaller batch-sizes incur higher inference overheads (e.g., weight loading) but can better overlap frontend and backend stages.
Furthermore, splitting queries into $n$ smaller batches can degrade quality as the top-$k$ items in each stage are set by stitching the top-$\frac{k}{n}$ items in each batch. 
Using \Infra, we ensure the system maintains high-quality and splits queries into four sub-batches for workloads studied in this paper.



\subsection{Customization of \Accel\ micro-architecture}
Below we detail \Accel's micro-architectural design space.

\textbf{Takeaway 5:} \textit{Splitting monolithic systolic arrays into sub-arrays improves recommendation inference throughput by concurrently processing multiple models and queries.}

As recommendation comprises large input working set sizes, \Accel~implements a weight stationary, systolic array-based MLP engine~\cite{samajdar2018scale, tpu, eyeriss}.
To concurrently process multiple stages and queries, \Accel~dynamically splits a monolithic array into independent sub-arrays~\cite{planaria}.  
Figure~\ref{fig:rpaccel_dse}(a) illustrates the benefit of a reconfigurable systolic array for multi-stage recommendation. 
We show the MAC utilization for various array sizes and models.
Larger arrays achieve lower latency but suffer from lower utilization when processing small models (i.e., RM$_{\textrm{small}}$). 
In fact, when processing a two-stage pipeline, the fixed, monolithic array has an average utilization of only 30\%, as it is overprovisioned for the frontend (i.e., RM$_{\textrm{small}}$ ranking 4K items).
Splitting the monolithic array into smaller units improves utilization to 60\%, doubling throughput at comparable latency.

\begin{figure}[t!]
  \centering
  \includegraphics[width=\columnwidth]{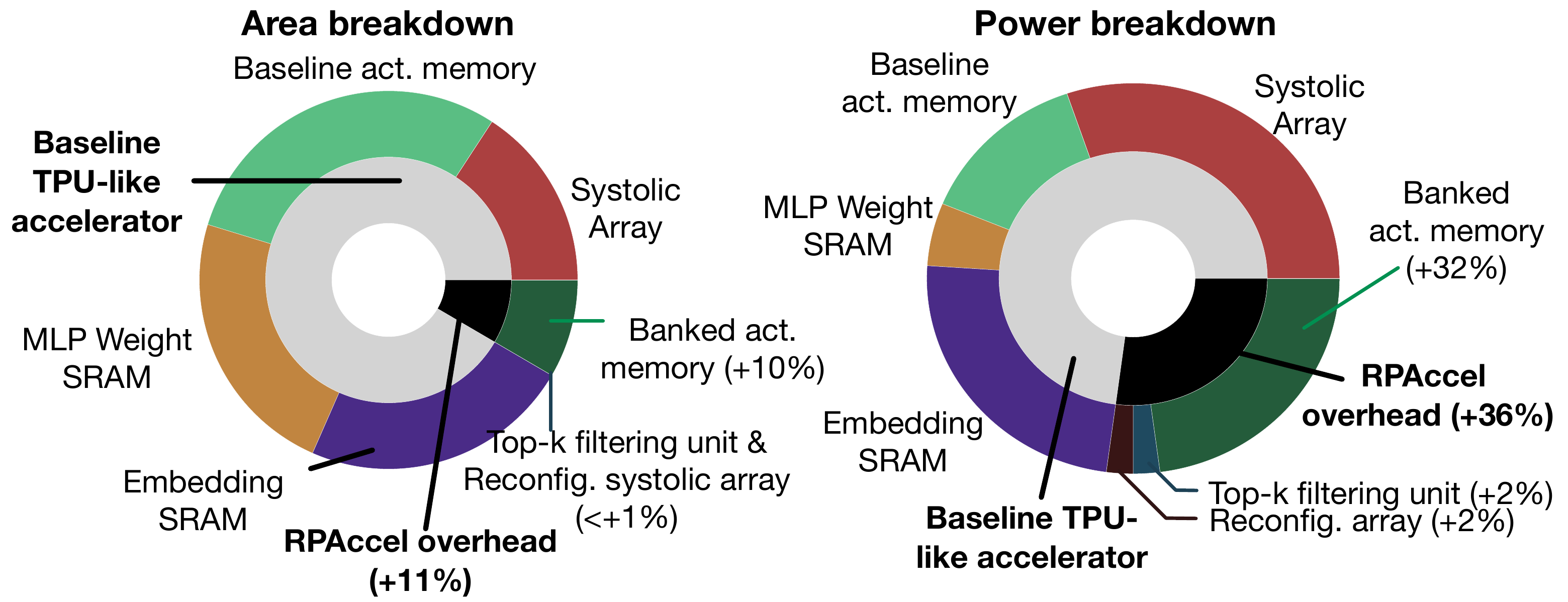}
  \vspace{-2.5em}
  \caption{Compared to the baseline, \Accel\ incurs 11\% and 36\% area (left) and power (right) overheads.}
  \label{fig:rpaccel_area}
  \vspace{-1em}
\end{figure}

Note, \Accel's reconfigurable systolic array is inspired by prior work which proposes a fission architecture to split monolithic arrays into sub-arrays for multi-tenancy~\cite{planaria}.
Customized for multi-stage recommendation, \Accel\ eliminates complex, omni-directional interconnects, incurring a lower area and power penalty (i.e., 13\% area and 21\% power in \cite{planaria} versus 6\% and 11\% in \Accel)\footnote{Following the baseline~\cite{planaria}, we exclude on-chip SRAM when comparing area and power. Figure~\ref{fig:rpaccel_area} includes SRAM overheads.}, and extends the reconfigurability in response to application QPS and SLA targets.

\textbf{Takeaway 6:} \textit{Implementing top-$k$ user-item filtering units in specialized hardware eliminates PCIe overheads.}

Based on the predicted CTR, top scoring user-item interactions must be filtered and forwarded to subsequent recommendation stages.
Prior recommendation accelerators only process MLP inference~\cite{centaur, jiang2021microrec}.
Thus, the filtering step is offloaded to host-processors incurring high PCIe overheads~\cite{centaur}.
To eliminate communication overheads, \Accel~implements a set of on-chip top-$k$ filtering units (see Figure~\ref{fig:MSRAccel} middle, blue).
One approach to identify the top-$k$ user-item pairs is to sort all CTR scores.
Unfortunately, sorting latency scales with the number of items to rank, potentially consuming tens-thousands of cycles for recommendation given large input sizes.
Furthermore, existing hardware sorting units consume significant area and power~\cite{hwsort}.

Instead, \Accel~exploits two unique properties of recommendation inference to simplify the filtering unit.
First, between stages, the final top-$k$ user-item pairs need not be ordered---\Accel\ implements an approximate, bucketing design.
Second, the final MLP layer produces one CTR score per cycle leading to a streaming filtering unit design.


Figure~\ref{fig:rpaccel_dse}(b) shows the resulting top-$k$ filtering unit.
The filtering unit maintains $N$ bins (e.g., $N$=16).
Each bin represents user-item pairs of a specific CTR score range between 0 and 1.
As a new CTR score arrives every cycle, the filtering unit adds the user-item $id$ to the corresponding bin and increments its counter.
For example, Figure~\ref{fig:rpaccel_dse}(b) shows the top bin counts user-item pairs with CTR scores between 0.9 and 1 (high quality).
Based on the CTR score, the user-item pair $id$ is stored in a dedicated portion of the systolic array banked weight SRAM.
Storing all (4K) user-item $id$ pairs consumes 12\% of the weight SRAM.
To reduce this overhead, \Accel\ skips user-item pairs with low CTRs.
Using \Infra, we set a minimum CTR threshold of $0.5$, reducing the overhead on weight memory to 3\%.
Once all user-item CTRs are categorized, the filtering unit identifies and copies at least top-$k$ user-item pairs indicated from the highest $n$ bins to DRAM.
These $id$s uniquely reference continuous and categorical inputs for subsequent stages.

Given the streaming design, the performance overhead of the filtering step is set by the latency it takes to identify and send user-item $id$s from the highest bins to main memory.  
We find this takes a couple hundred accelerator cycles, negligible compared to model inference.
Although each sub-array in \Accel's reconfigurable systolic array requires a separate top-$k$ filtering unit, the area and power overheads are small (see Figure~\ref{fig:rpaccel_area}) and there is no degradation in quality.

\textbf{Takeaway 7:} \textit{Asymmetrically-provisioned embedding caches tailored for each of the multi-stage models minimizes memory access latency.}

Recent work shows embedding table operations suffer from irregular memory access patterns, low compute intensity, and high storage capacities~\cite{gupta2020architectural}.
Consequently, the performance of embedding table operations is bounded by embedding vector fetch latency.
Prior work exploits the power-law distribution of embedding lookups to cache frequently accessed vectors on-chip~\cite{adnan2021high, ke2020recnmp, centaur}.
The embedding caches in prior work however assume a single stage recommendation model.

Instead, \Accel~implements an embedding cache customized for multi-stage recommendation by comprising (1) a \textit{static embedding cache} that is  provisioned statically for hot embedding vectors from both frontend and backend stages, (2) a \textit{look-ahead embedding cache} that stores embedding vectors for in-flight queries. It also prefetches lookups for later stages in \Accel's pipeline optimization (Figure 9(right)).
As shown in Figure~\ref{fig:MSRAccel}(left), input embedding IDs arrive either from the host processor for frontend models or the output of top-$k$ filtering units for backend models.
Based on the IDs, the embedding gather unit first checks if the corresponding vectors are in the caches.
If yes, the embedding vectors are returned to the ``Dense-input SRAM'' to be processed by the MLP-top layers.
If not, the embedding gather unit retrieves the vectors from DRAM to the look-ahead embedding cache.

\textbf{Embedding cache provisioning.}
Following data center AI accelerators with 24MB capacity~\cite{tpu}, we start with 16MB for embedding caches (8MB in MLP weights/activations).
The size of the \textit{look-ahead} cache is bounded by the number of items ranked in backend stages, size of embedding vectors, and maximum number of queries in flight.
For the worst case we conservatively provision 4MB for the  \textit{look-ahead} cache. 
This leaves 12MB for the \textit{static embedding} cache.
Figure~\ref{fig:rpaccel_dse}(c) shows the impact of asymmetrically provisioning memory for frontend and backend models on the average memory access time (AMAT) for embeddings. 
With a 128 byte cache line size, the embedding vector size of $RM_{\textrm{large}}$, we find the fraction of storage devoted to the frontend versus backend depends on the item filtering ratio between stages.
Given a filtering ratio of one-eighth for Criteo, we provision equal memory capacity for the frontend and backend.

\textbf{Area and power breakdown.} Figure~\ref{fig:rpaccel_area} illustrates the area and power overheads of the proposed optimizations compared to the baseline, TPU-like recommendation accelerator~\cite{centaur}.
The combination of the reconfigurable MLP unit, top-$k$ filtering unit, and multi-stage aware embedding cache incurs a total of 11\% area and 36\% power overhead, moderate compared to \Accel's performance benefits (see Section~\ref{sec:accel_eval}).

\section{Evaluation of \Accel~At-Scale}~\label{sec:accel_eval}
In this section we evaluate the performance of \Accel\ at-scale.
Instrumenting \Infra\ with the simulated \Accel\ we study the proposed hardware solutions in terms of quality, tail-latency, and system-throughput.
We study \Accel\ using publicly available models and datasets; and also project the quality and performance trends for future recommendations.


\begin{figure}[t!]
  \centering
  \includegraphics[width=0.95\columnwidth]{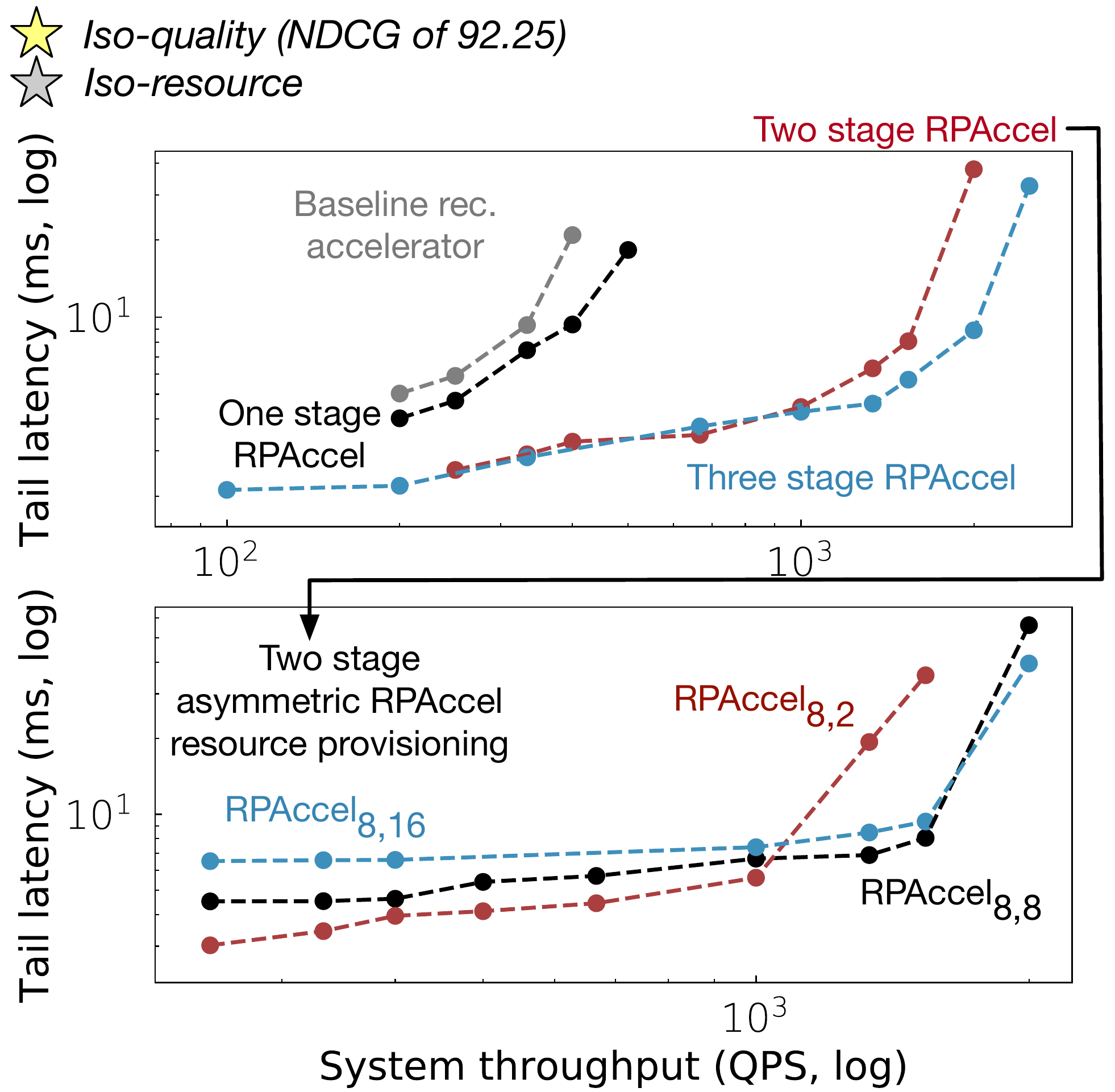}
  \vspace{-1em}
  \caption{(Top) At iso-quality and hardware resources, co-designing multi-stage models with hardware enables lower tail-latency and higher system throughput.
  (Bottom) Asymmetrically provisioning \Accel\ resources across stages further improves performance.}
  \label{fig:accel_analysis}
  \vspace{-1em}
\end{figure}

\subsection{\Accel\ evaluation on open-source use cases}
\textbf{Takeaway 8:} \textit{By accelerating multi-stage recommendation, \Accel\ achieves 3$\times$ lower latency and 6$\times$ higher throughput compared to baseline, single-stage designs.}

Given fixed hardware resources, Figure~\ref{fig:accel_analysis}(top) illustrates the tradeoff between throughput and latency as we vary the \Accel-provisioning decisions for all stages.
The baseline follows Centaur~\cite{centaur}---a single-stage recommendation accelerator which implements a TPU-like systolic array~\cite{tpu} and uses the host-processor to filter top-$k$ interactions. 
The baseline achieves a 6$ms$ and 21$ms$ tail-latency at the inference throughput of 200 and 400 QPS, respectively.
While achieving the same quality, the single-stage \Accel~design achieves a 4.5$ms$ and 9$ms$ tail-latency at 200 and 400 QPS, respectively.
Furthermore, decomposing recommendation into finer-grained pipeline enables a minimum latency of 2.1$ms$ at 200 QPS or, at 6$ms$ a throughput of 1300 QPS--- 3$\times$ and 6$\times$ improvement over the single-stage baseline, respectively.
The latency reduction and throughput increase owe \Accel's software and hardware optimizations.

\textbf{Takeaway 9:} \textit{Asymmetrically provisioning accelerator based on multi-stage recommendation models resources unlocks lower tail-latency and higher system-throughput.}

Figure~\ref{fig:accel_analysis} (bottom) illustrates the benefit of asymmetrically provisioning \Accel\ resources across stages.
For a two-stage recommendation pipeline, the frontend is fixed with sub-arrays while the backend implements two (i.e., \Accel$_{8,2}$), eight (i.e., \Accel$_{8,8}$), and sixteen sub-arrays (i.e., \Accel$_{8,16}$).
All experiments assume iso-hardware resources while achieving the maximum quality target (i.e., NDCG of 92.25).

\begin{figure}[t!]
  \centering
  \includegraphics[width=0.95\columnwidth]{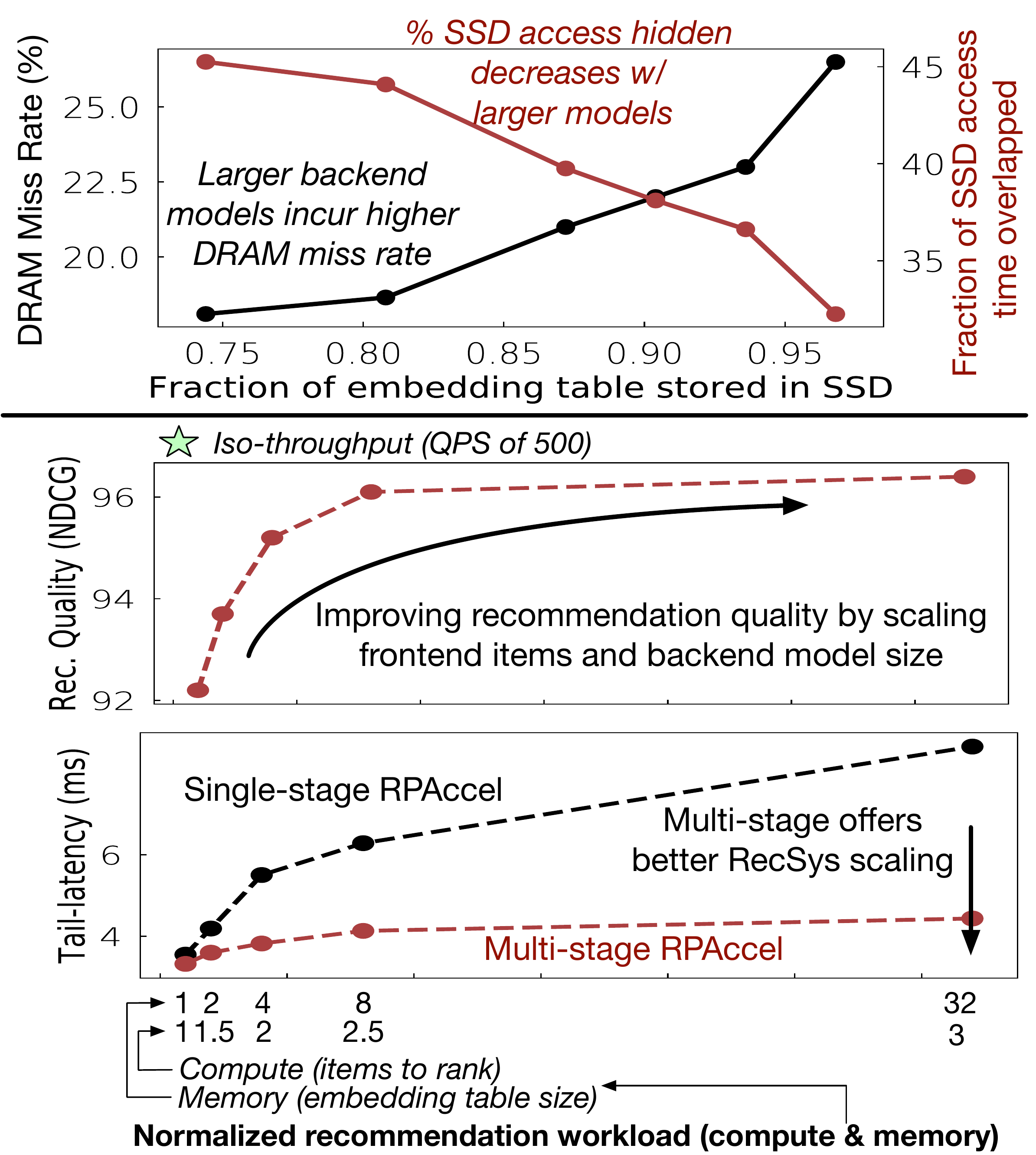}
  \vspace{-0.5em}
  \caption{(Top) Projecting the performance impact of scaling recommendation models to higher capacities requiring SSD storage.
  (Bottom) Compared to the single-stage accelerator baseline, \Accel\ provides graceful performance trends with future model sizes by also scaling items to rank to overlap frontend and backend stages. }
  \label{fig:future}
  \vspace{-1em}
\end{figure}

Compared with the homogeneous accelerator (i.e., \Accel$_{8,8}$),  aggregating the backend into fewer, larger arrays \Accel$_{8,2}$ reduces the latency at low throughput by 1.5$\times$.
Similarly, at high system load, splitting the backend into multiple, smaller units \Accel$_{8,16}$ reduces the latency by 1.4$\times$.
Given application-level latency and system targets, asymmetrically provisioning \Accel~resources across stages widens the design space of recommendation services.
Building on prior art, \Accel~resources are dynamically reconfigured to meet varying targets, given workload demands~\cite{planaria, kwon2019herald, song2019hypar}.

\subsection{\Accel~evaluation on future models}
So far we have analyzed the performance of \Accel\ on open-source use-cases.
However, recent literature shows production-scale recommendation models are rapidly growing in size, outpacing DRAM capacity and even reaching TBs in size~\cite{lui2020understanding}.
One promising path to enabling future, production-scale models is to use higher capacity memories such as SSDs~\cite{wilkening2021recssd,eisenman2018bandana}.
Here we consider the performance implications of SSDs on \Accel.

Storing larger embedding tables in SSD lowers embedding locality and  degrades performance.
Figure~\ref{fig:future}(top) shows the impact of larger embedding tables on embedding locality.
While frequently accessed embedding vectors are stored DRAM, a larger portion of these tables are stored in the SSD (x-axis).
For example, increasing the size of RM$_{\textrm{large}}$ by 32$\times$ requires storing 97\% of the embedding tables in SSD.
This also causes increases DRAM miss rates from 17\% to 28\%.
Recall, \Accel~pipelines frontend and backend stages---allowing the accelerator to overlap long latency SSD accesses in the backend.
However, Figure~\ref{fig:future}(top) shows with growing embedding table sizes, a smaller fraction of the accesses can be overlapped causing an increase in latency.


\textbf{Takeaway 10:} \textit{Compared to baseline single-stage accelerators, \Accel\ achieves higher quality and performance when scaling both frontend and backend stages towards future recommendation engines.}

In addition to scaling embedding tables in backend models (e.g., model size), one can also increase the number of items to rank in the frontend (e.g., compute demand).
Figure~\ref{fig:future}(bottom) shows the impact of scaling both frontend and backend stages (x-axis) on quality.
Starting from the baseline configuration, we project increasing model size by 32$\times$ and compute complexity from ranking 4K items to 12K items improves quality from an NDCG of 92.25 to 96.

\begin{figure*}[t!]
  \centering
  \includegraphics[width=\textwidth]{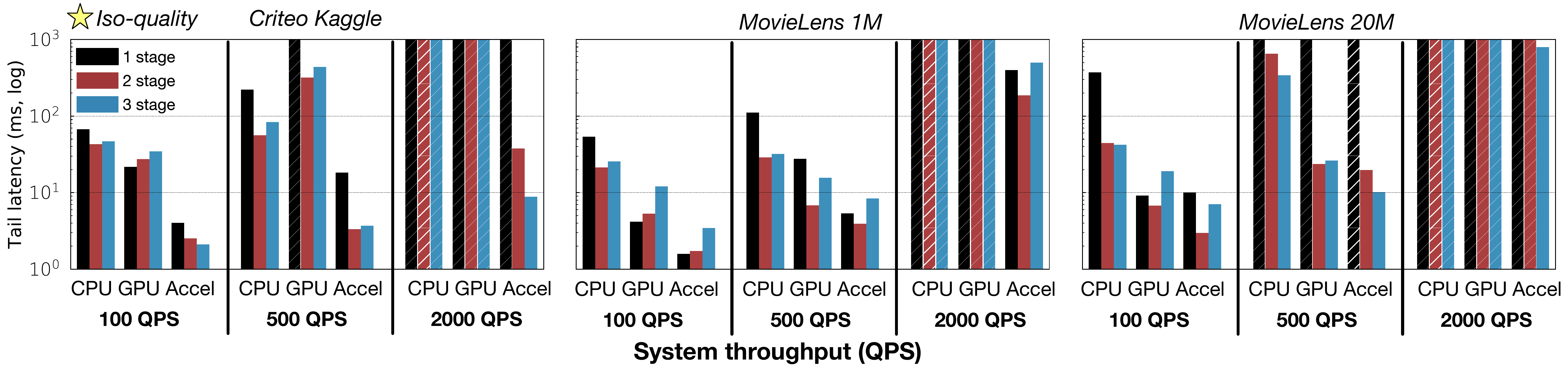}
  \vspace{-2em}
  \caption{Summary of RecPipe results at iso-quality for the Criteo, and MovieLens 1M and 20M datasets. 
  For each dataset, we show the tail-latency (log scale) for three system loads and hardware platforms.
    Configurations are greyed out when system load is not met.
  The optimal multi-stage design varies across loads, hardware platforms, and datasets.} 
  \label{fig:summary}
  \vspace{-1em}
\end{figure*}

Increasing the items to rank allows \Accel~to more effectively overlap the frontend and backend stages.
Figure~\ref{fig:future}(bottom) shows the corresponding tail-latency impact on scaling compute and memory complexity assuming iso-throughput (QPS of 500).
We show two configurations: single-stage (black) and multi-stage (red) \Accel.
By overlapping frontend and backend stages, the multi-stage design achieves higher performance for larger recommendation engines compared to the single-stage design.
More generally, we show the importance of tightly-coupling algorithm and hardware scaling for future recommendation engines; \Infra\ and \Accel\ open such new co-design opportunities. 

\section{Summary of \Name\ Results}~\label{sec:summary}
Figure~\ref{fig:summary} summarizes the performance benefits of the proposed solutions, co-designing models and hardware for multi-stage recommendation.
The results show the tail-latency across three datasets (i.e., Criteo Kaggle, MovieLens 1M and 20M~\cite{criteo,movielens1m}), system loads (i.e., QPS of 100, 500, 2000), and hardware platforms (i.e., CPU, GPU, Accel). 
The colored bars distinguish between one- (black), two- (red), and three- (blue) stage recommendation pipelines.
Following our previous analysis, CPU designs assume CPU-only execution (Section~\ref{sec:cpueval}).
For GPU-based configurations, 1-stage designs represent GPU-only execution; 2-stage and 3-stage designs represent heterogeneous GPU-CPU execution (Section~\ref{sec:gpu}).
\textit{Accel} configurations assumes \Accel-only execution (Section~\ref{sec:accel}-\ref{sec:accel_eval}).
Across the system loads and datasets, \Infra~reduces tail-latency by an average of 3.2$\times$ on commodity hardware; compared to prior recommendation accelerators, \Accel~reduces tail-latency by 4.3$\times$ on average.



\textbf{Differences across system loads.} Across  different system loads, the optimal multi-stage configuration and hardware platform varies. 
For instance, with the Criteo dataset on GPU-enabled hardware, between low (QPS of 100) and medium (QPS of 500) loads, the optimal number of stages varies from one to two.
Similarly, for Criteo, the optimal hardware backend between low and medium loads changes from GPUs to CPUs, respectively.
Differences across system loads owe to varying system optimization strategies for maximizing throughput versus minimizing latency; for example, throughput is maximized by processing multiple queries concurrently while latency is minimized by accelerating individual queries.


\textbf{Differences across datasets.} In addition to varying system loads, the optimal multi-stage configuration and hardware platform varies across datasets. 
For example with commodity hardware, on the Criteo dataset, CPUs achieve lower tail-latency than GPUs for system loads above 100 QPS; on the other hand, GPU-based designs outperform CPU-only execution for both MovieLens datasets.
With \Accel, tail-latency is optimized with the deeper three-stage pipeline for MovieLens-20M at 500 and 2000 QPS and all loads for Criteo; on MovieLens-1M however two-stage is optimal.
Differences across datasets owe to the Criteo implementing DLRM~\cite{naumov2019deep} with higher embedding capacities while MovieLens implementing neural matrix factorization models dominated by MLP layers; furthermore, across stages the number of items to rank reduces by roughly 5$\times$, 2.5$\times$, and 4$\times$, on Criteo, MovieLens 1M, and MovieLens 20M, respectively.
These differences highlight the need to co-design multi-stage recommendation parameters with the underlying hardware early in the design process using frameworks like \Infra.


\textbf{Benefits of \Accel.} 
Compared with CPUs and GPUs, \Accel~significantly reduces tail-latency of multi-stage recommendation across different datasets and system loads.
In fact, in many cases (e.g., Criteo and MovieLens20M datasets) \Accel~is optimized with deeper pipelines compared to commodity GPUs;
This is a direct result of extracting data-level and model-level parallelism opportunities across multi-stage recommendation and eliminating high-overhead host-accelerator communications that \Accel\ enables.




\section{Related Work}
While systems and computer architecture researchers have proposed various solutions to optimize cloud-scale personalized recommendation models, relatively little work explores co-design opportunities between models and hardware to jointly optimize quality and performance, as well as the unique characteristics of multi-stage recommendation.

\textbf{DNN-based recommendation models.}
To improve content personalization, recommendation models are growing rapidly in size and complexity~\cite{lui2020understanding, dienzhou2019deep, dinzhou2018deep, mtwnd}. 
Tackling the growing model sizes, researchers have proposed techniques to compress embedding tables while preserving accuracy~\cite{ginart2019mixed, shi2020compositional,ghaemmaghami2020training,ttrec}.
Alternatively, one can decompose large monolithic models into multi-stage pipelines.
Industry publications show multi-stage designs are used for serving content on Youtube~\cite{mtwnd} and Instagram~\cite{instagramMsr, godaStackingTwitteryRecSys}.
To balance recommendation quality and model complexity, machine learning researchers have explored a variety of modeling techniques to train each stage of the multi-stage pipeline~\cite{kang2019candidate}.
However, in prior work, the multi-stage recommendation systems are designed to maximize quality, independent on the underlying hardware.
\Name~extends prior art by co-designing the multi-stage models and underlying hardware---commodity and specialized---in order to tightly co-optimize quality, tail-latency, and throughput for data center scale deployment.


\textbf{Specialized recommendation hardware.} 
Lots of research effort has been devoted to design specialized hardware for deep learning---especially MLPs, CNNs, and RNNs~\cite{minerva, eie, eyeriss, gupta2019masr, pentecost2019maxnvm,samajdar2018scale, tpu, caulfield2016cloud, zhangcompact, dadiannao, cambricon, epur, choi2020prema}. 
However, recommendation systems pose distinct challenges owing to their network architectures and use cases~\cite{gupta2020architectural, hsia2020cross, naumov2019deep}.
Given its importance, hardware proposals for accelerating recommendation models have begun to emerge~\cite{gupta2020deeprecsys,kwon2019tensordimm,ke2020recnmp,centaur,kwon2020tensor,naumov2020deep,acun2020understanding, jiang2021microrec, sambanova, neuchips, asgari2021fafnir, xie2020kraken, kim2020trim}.
While prior work focuses on improving hardware efficiency given fixed workloads, \Name\ brings quality into the mix. 
Accounting for both quality and performance, this work co-designs multi-stage models and hardware.
In addition to \Infra's post-training inference scheduler on commodity hardware, we compare the proposed \Accel\ to a state-of-the-art TPU-like baseline recommendation accelerator, Centaur~\cite{centaur}; compared to the baseline, we demonstrate that by co-designing models and hardware, \Accel\ jointly improves recommendation quality, tail-latency, and throughput.
\section{Conclusion}
Given the growing prevalence of personalized recommendation, architects have invested significant resources improving recommendation inference efficiency.
While proposed solutions tackle different compute and memory bottlenecks, they do not directly co-optimize quality and performance.
In this work we propose \Name, a system for co-designing models and hardware to jointly optimize quality, tail-latency, and throughput.
First, \Infra\ splits monolithic models into multi-stage pipelines exposing unique system optimization opportunities.
Next, we design an inference scheduler that maps multi-stage recommendation across CPUs and GPUs.
Finally, we deign a novel hardware accelerator for multi-stage recommendation which achieves high-quality while improving latency and throughput by up to 3$\times$ and 6$\times$, respectively, over a baseline TPU-like recommendation accelerator.


\bibliographystyle{IEEEtranS}
\bibliography{references}


\end{document}